\documentclass[11pt,drafunctiontcls,onecolumn]{IEEEtran}
\usepackage{./mylay,epsfig}
\usepackage{amsmath}
\usepackage{epsfig}
\usepackage{bbm}
\usepackage{tikz}
\usepackage{graphicx,caption}
\usepackage{subfig}
\usepackage{bbm}
\usepackage{algorithmicx}
\usepackage[ruled]{algorithm2e}
\usepackage{cite}

\newcommand{\com}{Y_0}

\newcommand{\comt}{\tilde{Y}_{0}}

\newcommand{\pq}{p_{{}_{q}}}
\newcommand{\pqt}{p_{{}_{\tilde{q}}}}
\newcommand{\p}{p}

\newcommand{\w}{{\bf w}}
\newcommand{\s}{{\bf s}}

\IEEEoverridecommandlockouts
\begin{document}
\sloppy
\title{Community Detection Using Slow Mixing Markov Models} 
\author{Ramezan~Paravi~Torghabeh~~\IEEEmembership{Member,~IEEE,}
        and~Narayana~Prasad~Santhanam~\IEEEmembership{Member,~IEEE}
\thanks{The authors are with the Department
of Electrical Engineering, University of Hawai`i, M\={a}noa (Email:$\{$paravi, nsanthan$\}$@hawaii.edu).}
}
\maketitle

\begin{abstract}
  The task of \emph{community detection} in a graph formalizes the
  intuitive task of grouping together subsets of vertices such that
  vertices within clusters are connected tighter than those in
  disparate clusters. This paper approaches community detection in
  graphs by constructing Markov random walks on the graphs. The mixing
  properties of the random walk are then used to identify
  communities. We use coupling from the past as an algorithmic
  primitive to translate the mixing properties of the walk into
  revealing the community structure of the graph. We analyze the
  performance of our algorithms on specific graph structures,
  including the stochastic block models (SBM) and LFR random graphs.
\end{abstract}

\begin{keywords}
\emph{Random Walk, Slow Mixing Markov Processes, Graph Partitioning, Community
Detection, Correlation Clustering, Coupling From the Past.} 
\end{keywords}

\section{Introduction}
Applications in biology~\cite{ben1999clustering, girvan2002community,jonsson2006cluster}, social sciences~\cite{handcock2007model} and recommendation systems~\cite{mirza2003studying,shepitsen2008personalized} can be abstracted into
\emph{community detection} problems, where one attempts to find
subsets of nodes that share common
properties~\cite{fortunato2010community}. Typically, one constructs a
graph with vertices corresponding to the objects to be organized
together, while the graph edges represent connections between them. In the
simplest setting for the community detection problem, one would
partition the vertices into clusters that are strongly connected
within themselves, while the cross- cluster connections are weak. One
could consider more nuanced extensions as well, where vertices could
belong to multiple clusters, or where edges may be weighted.

To evaluate community detection algorithms, several variants of random
graph models play an important role in generating test graphs that
capture properties of real
networks~\cite{erdds1959random,erdos1960evolution,newman2002random} in
some way. One widely used model is the \emph{Stochastic Block Models}
(SBM). In this case, information theoretic limits for exact and partial
recovery has been addressed as well 
in~\cite{abbe2014exact,mossel2014reconstruction,abbe2015community,massoulie2014community}. 
For community detection in SBMs, the approaches that have been widely considered are the maximum likelihood
approach, semidefinite programming and belief propagation
methods~\cite{hajek2014achieving,hajek2015achieving,decelle2011asymptotic}.

Many approaches for community detection implicitly require that the
true number of clusters be known \emph{a-priori}.  An alternate
approach is taken in \emph{Correlation Clustering} where an objective
function is designed so optimizing the objective function cost will
automatically capture the underlying partition~\cite{Bansal04}. The
correlation clustering problem is often formalized as an integer linear
program. In \cite{AilonCN05}, a randomized expected $3$-approximation
algorithm for correlation clustering was provided.

In biological networks, one of the core algorithms used traces back
to~\cite{girvan2002community,newman2004finding} and is based on what
is called \emph{centrality indices}. The idea is to remove
cross-cluster edges--if there are a few tight communities connected
sparsely by inter-cluster edges, most cross-community shortest paths
will go through the cross-cluster edges. The algorithm formalizes this
insight and removes what it thinks are cross-cluster edges, leaving
behind the communities. 
\ignore{The modularity-based approaches
\cite{chen2009detecting,newman2006modularity} are used for this
problem as a tool for identifying structure of the underlying
graph. The modularity measure evaluates the quality of a given graph
partition by comparing the given partition with a randomized version
of the graph in which the degrees are preserved.}

The closest other line of work to our approach is spectral
clustering~\cite{newman2006finding,white2005spectral,meilpa2001learning,kannan2004clusterings}. This method uses a given pairwise similarity between nodes in a network to assign weights to the edges, and uses the eigenvectors
of the (symmetric, normalized) graph Laplacian to assign nodes to
clusters.  For two clusters, the Fiedler eigenvector can be partitioned
around the median to yield clusters and for more than two clusters,
similar extensions can be used involving $k$ eigenvectors. While the
implementation is simple, a couple of drawbacks need to be
addressed---the pairwise similarity is often not automatically given
in community detection problems, the Laplacian could be ill
conditioned and the number of clusters need to be known in advance, see \eg~\cite{von2007tutorial,ng2002spectral}. 

\subsection*{Markov Processes}
Let $\p$ be an
irreducible and aperiodic first-order Markov process
$\{ Y_{i}\}_{i=-\infty}^{\infty}$ taking values in a finite alphabet
$\mathcal{S}$ with $|\mathcal{S}|=r$. Let $Q=[p(\s|\s')]$ with
$\s,\s'\in\mathcal{S}$ denote the transition probability matrix of the
process. Here, $p(\s|\s')\triangleq \p (Y_1=\s|Y_0=\s')$ denotes the transition probability from
$\s'$ to $\s$. Furthermore, let $\pi$ denote the unique stationary
distribution of $p$, the unique solution of $\pi Q=\pi$. 
Namely, $\pi$ is the unique solution to
\begin{equation}
\label{mar_stat}
\pi(\s)=\sum\limits_{\s' \in \mathcal{S}}\pi(\s') p(\s|\s'),
\end{equation}
for all $\s \in \mathcal{S}$. Clearly $\p (Y_1=\s)=\pi(\s)$.

\subsection*{Core idea} Empirical properties of finite sized samples
from Markov processes need not reflect staionary properties---\eg the
number of occurances of a string $\w$ in a finite sized sample from a
Markov process $p$ need to yield a good estimate of $p(\w)$. Roughly
speaking, when the empirics of the sample eventually reflect the
stationary properties, we say that the process has mixed. The core idea
in this paper revolves around interpretation of the random walk before
mixing has occured. Naturally then, to build clustering and community
detection algorithms on graphs, we build Markov random walks such that
the restriction of the random walk to within any one cluster mixes much
faster than the overall walk itself. Once we obtain a walk like that, a
careful interpretation of samples from the random walk along the lines of
the theorems we obtain in~\cite{asadi2014stationary} should reveal the community structure
of the graph.

\section{Community Detection using Slow Mixing Markov Processes}
In order to identify communities in the graph, we define random walk
on the graph, such that the mixing properties of the walk reveal the
clusters. To do so, we use Coupling From The Past (CFTP)~\cite{PW96}
to sample appropriately from the constructed random walk. CFTP was
devised by Prop and Wilson and allows to obtain samples distributed
according to the \emph{exact} stationary distribution of a Markov
process. In CFTP, the idea is to simulate different copies of a given
Markov chain, each starting from a different initial state, but in a
way that is often described as ``backwards'' in time---details will
be outlined in Section~\ref{s:pcftp}. While each chain
is marginally faithful to the same Markov law, different copies are
not independent from each other. Rather, the joint evolution is set up
so that chains \emph{coalesce}---once chains hit the same state at the same
time, their future evolution is identical. When all copies of the
chain coalesce at time $0$, the obtained sample is distributed
\emph{perfectly} according to stationary distribution.

Rather than sampling from the stationary distribution, we adapt the
CFTP approach to identify clusters before the chains coalesce. To
achieve this, we will identify subsets of state space which coalesce
together faster. In following section, we introduce the notion of
restriction of a Markov process to a subset $G \subseteq \mathcal{S}$
which will be translated into algorithmic rules for community
detection in graphs.

\subsection{Restriction of a Markov Process to $G$}
Let $G$ be an arbitrary nonempty subset of the state space $\mathcal{S}$
of the Markov process $p$ with $|G|=r'$. Let
$\{\tilde{Y}_{j}\}=\{\cdots, \tilde{Y}_{-1}, \tilde{Y}_0, \tilde{Y}_1,
\cdots\}$
be the restriction of $\p$ to $G$. Note that $\tilde{Y}_j \in G$, for
all $j$. We will refer to restricted process at $\tilde{p}$. By the strong Markov property, $\tilde{p}$ is also a
Markov process over $G$. Furthermore, if $\tilde{p}$ is aperiodic, then it has a stationary distribution which will be denoted by $\tilde{\pi}$. Let $\pi(G)$ denote the stationary distribution of $G$, i.e.
\[
\pi(G)=\sum\limits_{\w \in G} \pi(\w).
\]
The transition probability matrix of the process, $Q$, can be written as
\begin{equation}
\label{Rearranged}
Q =  \begin{pmatrix}
			Q_{GG} & Q_{GB} \\
             Q_{BG} & Q_{BB} 
     \end{pmatrix},
\end{equation}
where $B=\mathcal{S}\backslash G$. Note that $Q_{GG}$ and $Q_{BB}$ are square matrices of size $r' \times r'$ and $(r-r') \times (r-r')$, respectively. Before proceeding further, we prove following useful lemma.
\bLemma 
\label{mat_series}
Let $A$ be an $m \times m$ matrix with $\max\limits_{1\leq i \leq m} |\lambda_{i}|<1$ where $\lambda_{i}$'s are the eigenvalues of $A$. Then,
\[
(I-A)^{-1}=I+A+A^2+A^3+ \cdots 
\]
\Proof The condition $\max\limits_{1\leq i \leq m} |\lambda_{i}|<1$ insures that the infinite series $I+A+A^2+A^3+ \cdots $ is convergent. The equality follows by direct multiplication. 
\eLemma

\bTheorem The transition probability matrix of the Markov process $\tilde{p}$ is given by
\label{prop_rest}
\begin{equation}
\label{restricted}
\tilde{Q}=Q_{GG}+Q_{GB} (I-Q_{BB})^{-1}Q_{BG},
\end{equation}
where $I$ is the identity matrix with size $r-r'$.
Furthermore, $\tilde{p}$ has a unique stationary distribution $\tilde{\pi}$ given by
\begin{equation}
\label{restricted_stat}
\tilde{\pi}(\w)=\frac{\pi(\w)}{\pi(G)},
\end{equation}
for all $\w \in G$.
\ignore{
\Proof Fix $\w,\w' \in G$. For convenience, we define following vectors obtained from $Q$
\begin{equation*}
V \triangleq [q(\s |\w')], \quad   \s\in B
\end{equation*} 
and
\begin{equation*}
W \triangleq [q(\w |\s)],  \quad \s\in B.
\end{equation*} 
Note that $V$ is the row vector containing the set of transition probabilities from $\w'$ to all states in $B$. Similarly, $W$ is the column vector containing the set of transition probabilities from all states in $B$ to $\w$. To compute $\tilde{q}(\w|\w')$, we have
\begin{eqnarray}
\tilde{q}(\w|\w')&=&\pqt(\tilde{Y}_1=\w|\tilde{Y}_0=\w') \nonumber \\
&=& \pq(Y_1=\w|Y_0=\w')  \nonumber \\
&+& \pq(Y_2=\w, Y_1 \in B|Y_0=\w') + \cdots \nonumber \\
&=& q(\w|\w')+ \sum\limits_{k=1}^{\infty}\pq(Y_{k+1}=\w, Y_1^k \in B^k|Y_0=\w') \nonumber \\
&=& q(\w|\w') + \sum\limits_{k=1}^{\infty} V Q_{BB}^{k-1} W \nonumber \\
&=& q(\w|\w') + V \big(\sum\limits_{k=0}^{\infty} Q_{BB}^{k} \big) W \nonumber \\
&=& q(\w|\w') + V \big( I-Q_{BB}\big)^{-1} W.
\label{trans}
\end{eqnarray}
It only remains to show that the $(I-Q_{BB})^{-1}$ in \eqref{trans} exists or equivalently, $\sum\limits_{k=0}^{\infty} Q_{BB}^{k}$ is convergent. By using Lemma \ref{mat_series}, the sufficient condition for convergence is that all the eigenvalues of $Q_{BB}$ (say $\lambda_{\ell}$'s, $1 \leq \ell \leq r-r'$) lie strictly inside the unit circle. To see this, first note that by irreducibility of $Q$, $Q_{BB}$ and all its minors can not be  stochastic matrices. Second, every diagonal element of $Q_{BB}$ must be strictly less than 1, otherwise if we start from the corresponding state, we always remain on that state which contradicts irreducibility. Hence, $Q_{BB}$ is a non-negative matrix which always can be dominated ---element-wise, by an \emph{irreducible}\footnote{A non-negative matrix $A_{m \times m}$ is called irreducible if for all possible partitions $J$ and $K$ of $\{1,2,\cdots,m\}$, there exist $j\in J$ and $k \in K$ such that $a_{jk} \ne 0$.} stochastic matrix $L$. Then, Perron-Frobenius theorem (see \eg \cite{Sen06}) guarantees that $|\lambda_{\ell}| \leq 1$.
Next, we claim that $|\lambda_{\ell}|<1$ for all $1 \leq \ell \leq r-r'$. If not, then there exists some ${\ell}'$ such that $|\lambda_{\ell'}|=1$. Since $0\leq Q_{BB}\leq L$ with $L$ irreducible and $|\lambda_{\ell'}|=1$, Perron-Frobenius theorem for irreducible matrices implies that $Q_{BB}=L$ which is clearly a contradiction. Arranging \eqref{trans} for every pair of states in $G$ into the matrix form, we have \eqref{restricted}.
Since $\pq$ is irreducible, $\pqt$ is irreducible as well. Also, since $G$ is finite, every state in $\pqt$ is positive recurrent. Hence, $\pqt$ has a unique stationary distribution (see \eg \cite{Ros06}). 
Let $T_\w^{m}$ with $m\geq 1$ denote the $m$'th recurrence time of the state $\w$ in $\{Y_i\}$. Note that $T_\w^{m}-T_\w^{m-1}$ are \textit{i.i.d.} with expected value $1/\pi(\w)=\pi(\mathcal{S})/\pi(\w)$ \cite{Ros06}. Similarly, let $\tilde{T}_\w^{n}$ with $n\geq 1$ denote the $n$'th recurrence time of the state $\w$ in $\{\tilde{Y}_j\}$. Observe that $\tilde{T}_\w^{n}-\tilde{T}_\w^{n-1}$ are \textit{i.i.d.} with expected value $\pi(G)/\pi(\w)$. Therefore, since the stationary distribution of $\w$ in $\pqt$ equals to the reciprocal of its mean recurrence time,  we have
\begin{equation*}
\tilde{\pi}(\w)=\frac{\pi(\w)}{\pi(G)}. \eqed
\end{equation*}}
\eTheorem

\subsection{Partial CFTP}
\label{s:pcftp}
In this section, we first briefly review \textsc{CFTP} algorithm and
then introduce the notion of partial coupling which is the core idea
to identify clusters in our proposed community detection
algorithm. Recall that $\mathcal{S}$ is the state space of the Markov
process $\p$ with $|\mathcal{S}|=r$. In \textsc{CFTP} algorithm, we
run coupled copies of $\p$, a different copy from every state in $\mathcal{S}$.
We use Dobrushin coupling~\cite{galves2008exponential} to couple the different chains.

To run the CFTP for 1 step, we start all coupled copies from time -1, and
run them for one step. To run it for the second step, we start coupled copies
of the Markov process from every state in $\mathcal{S}$ at time -2, and run
the chains for one step. Now we reuse the evolution from step -1 to 0. See~\cite{PW96}
for why we choose this peculiar way of running the chains backward in time.

We represent the samples of $\p$ starting from time
$-n$ with state $\s$ by $\{Y_{-i}^{(\s,-n)}\}$ where
superscript $\s$ denotes the initial state. 
Note $Y_{-n}^{(\s,-n)}=\s$. 

\bDefinition
\label{reg_coel} 
We say that the \textsc{CFTP} algorithm has coalesced in the
\emph{regular sense} in $n$ steps if $n$ is the smallest number
such that for all $\s , \s' \in \mathcal{S}$
\begin{equation*}
Y_{0}^{(\s,-n)}=Y_{0}^{(\s',-n)}.
\label{con: reg_coel}
\end{equation*}
Let $Y_{0}$ denote the element that all chains have coalesced to,
namely $Y_0=Y_0^{(\s-n)}$, $\s\in\mathcal{S}$. Note also that the
definition does \emph{not} imply that time 0 is the 
point at which the chains $Y_0^{(\s,-n)}$ first coalesce.
\eDefinition

We will refer to $Y_{0}$ in Definition~\ref{con: reg_coel} as the output of \textsc{CFTP} algorithm. 
\bTheorem
\label{T:CFTP} [From~\cite{PW96}]
Let $\com$ be the output of the \textsc{CFTP} algorithm. Then, for all $\s \in \mathcal{S}$ 
\[
\mathbb{P}(\com=\s)=\pi(\s),
\] 
where $\mathbb{P}$ and $\pi$ denote the coupling distribution and the stationary distribution of $\p$, respectively.
\eTheorem

Let $G\subseteq \mathcal{S}$. For a fixed $n$ and $\w \in G$, we define
\begin{equation}
N_{-n}^{\w}\triangleq \sum\limits_{i=0}^{n-1}\mathbbm{1}(Y_{-i}^{(\w,-n)} \in G),
\label{num_occur}
\end{equation}
which denotes the number of occurrences of elements in $G$ along the realization of the copy of $\p$ starting from $\w$.   

\bDefinition
\label{partial_coel_stopping}
We say that the \textsc{CFTP} algorithm at time $-n$ is \emph{partially coalesced} with respect to $G$ if $-n$ is the smallest time index such that for all $\w, \w' \in G$
\begin{equation}
Y_{0}^{(\w,-n)}=Y_{0}^{(\w',-n)}, \text{ and } \, 
N_{-n}^{\w}=N_{-n}^{\w'}.
\end{equation}
Let $\comt=Y_0^{(\w,-n)}$ for all $\w\in G$. We refer to $\comt$ as the output of the algorithm.
\eDefinition

\ignore{
\bDefinition Define the \emph{coalescence time \textit{w.r.t.} $G$} as 
\begin{align*}
\tau_G \triangleq \min \big\{n \colon &\text{partial coalescence \textit{w.r.t.} $G$}\\ 
&\quad \text{ occures at time $-n$}  \big\}.
\end{align*}
\eDefinitionp
\bLemma 
\label{power_markov}
Let $\p$ be an aperiodic, irreducible Markov process taking values in
$\mathcal{S}$ with stationary distribution $\pi$ which is the unique
solution to $\pi=\pi Q$. If there exist a probability distribution
$\nu$ satisfying $\nu=\nu Q^k$ for some $k \in \mathbb{N}$, then
$\pi=\nu$.  \Proof Pick arbitrary $k \in \mathbb{N}$. Suppose
$\nu=\nu Q^k$. Note that $Q^k$ is a stochastic matrix which naturally
defines a Markov process taking values in $\mathcal{S}$. Such Markov
process is irreducible and positive recurrent, thus it has a unique
stationary distribution. By assumption, $\nu$ must be the unique
stationary distribution of the new process. On the other hand, $\pi$
satisfies $\pi=\pi Q^k$. Hence, $\pi=\nu$.  \eLemma }

\bTheorem 
\label{T:PCFTP-stopping}
Let $p$ be a Markov process over state space $\mathcal{S}$ and 
let $G$ be an arbitrary nonempty subset of $\mathcal{S}$. Assume that
the \textsc{CFTP} algorithm at time $-n$ has partially coalesced
\textit{w.r.t.} $G$ with $\comt \in \mathcal{S}$ as the output of the
\textsc{CFTP} algorithm per Definition~\ref{partial_coel_stopping}.
Suppose we further continue the simulation of the chain (starting from
$comt$), and let $j$ be the first time $Y_j\in G$ for some $j\ge1$. Let
$Y=Y_j$.
Then, for all $\w \in G$, we have
\[
\textbf{Pr}(\tilde{Y}=\w)=\frac{\pi(\w)}{\pi(G)},
\]
where $\pi$ denotes the stationary distribution of $\p$, respectively.
\eTheorem 

\ignore{
\bCorollary 
\label{cor:PCFTP-det}
Suppose there exists an arbitrary nonempty subset $G$ of $\mathcal{S}$ such that all the paths \emph{starting} from $\w \in G$ have been coalesced in the \emph{regular sense} at time $-n$. Furthermore, assume that partial coalescence \textit{w.r.t.} to $G$ happens in the interval $[-n, -n']$ for some $n'<n$. If for all  $\w,\w' \in G$
\[ N_{-n}^{\w}=N_{-n}^{\w'} \triangleq N \geq 1,
\]
where $N_{-n}^{\w}$ is defined in \eqref{num_occur}, then all the occurrences of common elements in $G$ within each path are distributed according to $\pi(\w)/\pi(G)$ as well.
\eCorollary
}
\subsection{Algorithm for Community Detection}
\label{cftp-community-det}
The mixing properties of the process can be used to identify communities in the set of states of the process. To this end, we need to consider finding a partition $\mathcal{C}=\{C_1, \cdots C_m\}$ of the state space $\mathcal{S}$ such that $C_i \subseteq \mathcal{S}$, $C_i\ne \emptyset$ for all $i$ and $C_i\cap C_j=\emptyset$ for $i\ne j$ with $\bigcup_{i=1}^m C_i=\mathcal{S}$. We will refer to individual $C_i$ as a {\bf{cluster}}.

Note that we can associate to every partition $\mathcal{C}$, a cost function $J(\mathcal{C})$ depending on the application. For instance, in the context of random walks on graphs, one convenient choice can be cluster editing cost as we will discuss in section \ref{sim: SBM}. The other parameter is the number of clusters $m$. In our approach, in contrast to traditional methods such as $k$-means, $m$ need not to be known a a-priori. In the Algorithm \ref{alg:comm-det}, the number of clusters will automatically will be determined by the algorithm and its value depend on the cost function $J$ we consider. The algorithm is based on Theorem \ref{T:PCFTP-stopping} which essentially describes how to identify a subset of states which have been partially coalesced. In essence, the algorithm identifies communities which have coalesced together during one sample run of \textsc{CFTP} algorithm. 

\textbf{Description of the Algorithm:} We consider one run of the \textsc{CFTP} algorithm and our setup is the same as original \textsc{CFTP}. We emphasize that the chains are simulated backward in time and the the random variables used during execution are reused. Recall that $|\mathcal{S}|=r$. Initially, we have a partition of state space which consists of all singletons, \ie  $\mathcal{C}_0=\{C_1, \cdots, C_r\}$ such that $C_i=\{ \s\}$ for some $\s \in \mathcal{S}$. As we proceed backward in time to obtain a sample from stationary distribution, we will identify a set of {\bf{critical times}} denoted by $T$, which are the times which some clusters have been coalesced.

\bDefinition During execution of \textsc{CFTP} algorithm, we say that a time index $n$ is a \emph{critical time} if there exist at least two clusters $C_1$ and $C_2$ such that for all $\w, \w' \in C_1 \cup C_2$ 
\begin{equation*}
Y_{0}^{(\w,-n)}=Y_{0}^{(\w',-n)} \, \text{ and } \, 
N_{-n}^{\w}=N_{-n}^{\w'},
\end{equation*}
where 
\begin{equation*}
N_{-n}^{\w}=\sum\limits_{i=0}^{n-1}\mathbbm{1} \bigg( Y_{-i}^{(\w,-n)} \in C_1 \cup C_2 \bigg).\eqed
\end{equation*}
\label{critical-time}
\eDefinitionp

Note that above definition can naturally be extended to more that two clusters. As a consequence, we can identify clusters which have been merged in critical times. Observe that when all the chains coalesce, we have a single partition consisting of all states. In other terms, we start off with all-singleton partition up to the point where we have a single partition. However, during this process, at critical times, we can identify clusters which merge together to yield a coarser partition. We can compute the cost associated to the particular partition that the sampling process yields and find the optimal partition $\mathcal{C}^*$ with respect to the cost function $J$. 

In following pseudo-code, we represent all-singleton partition by $\mathcal{C}_0$ and \textit{CFlag} denotes the coalescence flag. Here, we assume that we are minimizing a cost function $J$. A container $T$ is used to store the critical times.

\begin{algorithm}[!htbp]
  \textbf{Input}: A Markov process $\p$ over $\mathcal{S}$.\\
  \textbf{Output}: A partition $\mathcal{C}^*$ of $\mathcal{S}$.  \\
  \textbf{Initialize}: $k \leftarrow 1$, $\mathcal{C}=\mathcal{C}_0$, $\mathcal{C}^*=\mathcal{C}_0$, $T=[ 0]$, $CFlag=False$.\\
 \While{$CFlag=False$}{
 Simulate r copies of the chain starting at time $-k$. \\
 \If{$k$ is a critical time}
 {Update $\mathcal{C}$ by merging clusters which are coalesced;\\
 $T=T.Append \{k\}$;\\
 	\If{$J(\mathcal{C}) \leq J(\mathcal{C}^*)$} {$\mathcal{C}^*=\mathcal{C};$}
 }
 \eIf{All $r$ chains reach to some common state at time $0$}
 {$CFlag=True$;}
 {$k \leftarrow k+1$}
 } 
 \Return{$\mathcal{C}^*$}
 \caption{\textsc{Detecting Communities in a Markov process}}
 \label{alg:comm-det}
\end{algorithm}

\bRemark We need to carefully define the Markov process depending on the application on hand to exploit mixing properties of the process for revealing the underlying community structure. Furthermore, the choice of cost function depends on how the Markov process is defined. 
\eRemark

\bRemark Note that in general, the algorithm can stop at any critical time and still yields a partition (though perhaps not optimal). A criteria which empirically seems to be effective is to record the difference $\Delta T_i\triangleq T_i-T_{i-1}$ between consecutive critical times. If the difference exceeds a prescribed threshold, the algorithm will stop. As we discuss simulation results in next chapter, we will see that if the process is slow mixing, then the resulted clusters usually coincide with ground truth communities.   
\eRemark

\section{Simulation Results}
Our algorithm creates a random walk on
the nodes of the graph. We start different, coupled random walks from
different nodes. We then adapt the coupling from the past approach to
to identify clusters before the chain mixes (rather than sample from
the stationary distribution).

\textbf{Construction of Random Walk: } In order to use Algorithm \ref{alg:comm-det} for graph partitioning problem,
first we define a \emph{non-uniform} random walk on the graph as follows:

Given an undirected graph $G=(V,E)$ and $v \in V$, let $\mathcal{N}(v)$ be the set of neighbors of $v$ in $G$. We assign a weight to every edge $e=(v,w) \in E(G)$ given by
\begin{equation}
w(e)\triangleq w(v,u)=f\bigg(\big|\mathcal{N}(v) \cap \mathcal{N}(u)\big| \bigg),
\label{weighted_edgeweight}
\end{equation}
where $f(x)=x^r$ for some $r \in \mathbb{Z}^+$. Then, for all $u \in \mathcal{N}(v)$, the transition probability of the walk is given by
\begin{equation}
\mathbb{P}(u|v)=\frac{w(v,u)}{\sum\limits_{u' \in \mathcal{N}(v)}w(v,u')}.
\end{equation}
The above scaling can be interpreted how different similar neighbor nodes vote on transition probabilities. A similar scaling method (with exponential function) was used in \cite{meilpa2001learning} for image segmentation applications.
This reflects the fact that cross-cluster edges have smaller weight compare to in-cluster edges.

To evaluate the performance of the algorithm we considered random
graph generative models including the (i) Stochastic Block Models
(SBM), (ii) LFR models (random graphs with power law degree
distribution having well defined community structure
\cite{lancichinetti2008benchmark}), and (iii) real world networks with
known community structure are used to test the algorithm. The number
of communities is not known to our algorithm in advance, nor is any
generative model we may have used. Where relevant, we used cluster
edit distance (number of addition and deletion of edges needed to turn
a graph into disjoint cliques) and compared the results with the
state-of-the-art Correlation Clustering (CC-PIVOT)~\cite{AilonCN05} algorithm.

\subsection{Stochastic Block Model}
\label{sim: SBM}
Let $G(n,p,q,k)$ be the random graph model which will has following properties. Let $V=\{1,\cdots, n\}\triangleq[n]$ be the set of the nodes in $G$. Assume that there exists an underlying true partition $V=\bigcup_{i=1}^{k}V_i$ of the nodes such that $\sum_{i=1}^{k}|V_i|=n$. In this generative model, for all $w, v \in V$, if $v\in V_i$ and $w \in
V_j$ for some $i \neq j$, then $w$ is connected to $v$ with probability $q$. If $v, w\in V_i$ for some $i$, then $w$ is connected to $v$ with probability $p$. All choices of edges are independent from each other and $E(G)$ denote the set of edges in the $G$.

For SBM random graphs $G(n,p,q,k)$, we tested the performance of algorithm by varying model parameters $n$, $p$ and $q$ and number of communities in the graph. Note that given a partition of graph a $G$, we can associate a cost to that partition. One convenient choice is the number of edge addition and deletion in order to turn the graph into disjoint clusters. This measure is known as \emph{cluster editing} cost~\cite{Sham04}. This is in the spirit of \emph{correlation clustering} setting in which one tries to transform a given graph into disjoint cliques by using minimum number of edge deletion or edge addition~\cite{Bansal04}. 

In order to evaluate the quality of the recovered communities in the SBM, we first consider the average cluster editing cost. For random graphs with roughly equal size ground truth communities, we computed and compared the average cost of partitions obtained by our algorithm and \textsc{CC-PIVOT}. The simulation results are summarized in TABLE \ref{cost-compare}. The result show that the partitions found by our method are usually have smaller cluster editing cost. 

\begin{table}[!t]
\centering
\caption{Comparison of average cost of our algorithm and \textsc{CC-PIVOT} on $10$ samples taken uniformly at random from $G(n,p,q)$. All cluster have the same size while the number of clusters in the graph varies.}
\scalebox{0.8}{
\begin{tabular}{cccccc}
\hline \hline
  Cluster-Size  &  Num-Cluster  &  p  &  q   &  Our-Cost  &  CCPivot-Cost  \\
\hline \hline
	5        &      60       & 0.95 & 0.05 &   3020.3   &     3572.3     \\
	5        &      60       & 0.95 & 0.2  &   9803.4   &    11992.4     \\
	5        &      60       & 0.8  & 0.1  &   5249.1   &     6518.6     \\
	10       &      30       & 0.95 & 0.05 &   3382.5   &     4243.4     \\
	10       &      30       & 0.95 & 0.2  &  10425.2   &    12725.3     \\
    10       &      30       & 0.8  & 0.1  &   5748.9   &     7023.4     \\
	20       &      15       & 0.85 & 0.05 &   4059.8   &     5030.6     \\
	20       &      15       & 0.8  & 0.1  &    6646    &     8119.2     \\
	20       &      30       & 0.9 & 0.05 &  13977.6   &    17569.8     \\
 	20       &      30       & 0.9 & 0.2  &   41432    &    51590.5     \\
	20       &      30       & 0.8 & 0.2  &  41042.9   &    51225.3     \\
	20       &      30       & 0.7 & 0.2  &   40594    &    50626.8     \\
	30       &      20       & 0.8 & 0.1  &  24626.5   &    31612.9     \\
60       &      10       & 0.9 & 0.05 &  22302.9   &    22466.3     \\
 60       &      10       & 0.8 & 0.1  &  29711.2   &    36251.1     \\
       
       60       &      10       & 0.8 & 0.2  &  47510.2   &    57262.2     \\
       60       &      10       & 0.7 & 0.1  &  28827.9   &    35849.5     \\
       60       &      10       & 0.7 & 0.2  &   47028    &    55923.3     \\
       75       &       8       & 0.9 & 0.1  &  33773.6   &    37830.4     \\
       75       &       8       & 0.9 & 0.2  &  51431.9   &     59096      \\
       75       &       8       & 0.8 & 0.1  &  32683.4   &    38469.7     \\
       75       &       8       & 0.8 & 0.2  &  49895.8   &     58606      \\
       75       &       8       & 0.7 & 0.1  &  31452.8   &    38101.2     \\
       75       &       8       & 0.7 & 0.2  &  48284.7   &    57460.1     \\
      100       &       6       & 0.9 & 0.1  &  38898.6   &    39732.2     \\
      100       &       6       & 0.9 & 0.2  &  55986.7   &    61159.9     \\
      100       &       6       & 0.8 & 0.1  &  37517.8   &    40637.8     \\
      100       &       6       & 0.8 & 0.2  &  53975.2   &    61695.5     \\
      100       &       6       & 0.7 & 0.1  &  35612.8   &     41015      \\
      100       &       6       & 0.7 & 0.2  &  51948.8   &     60161      \\
\hline \hline
\end{tabular}}
\label{cost-compare}
\end{table}

\begin{figure}[h,scale=0.2]
  \centering
  \subfloat[][]{\includegraphics[width=.25\textwidth]{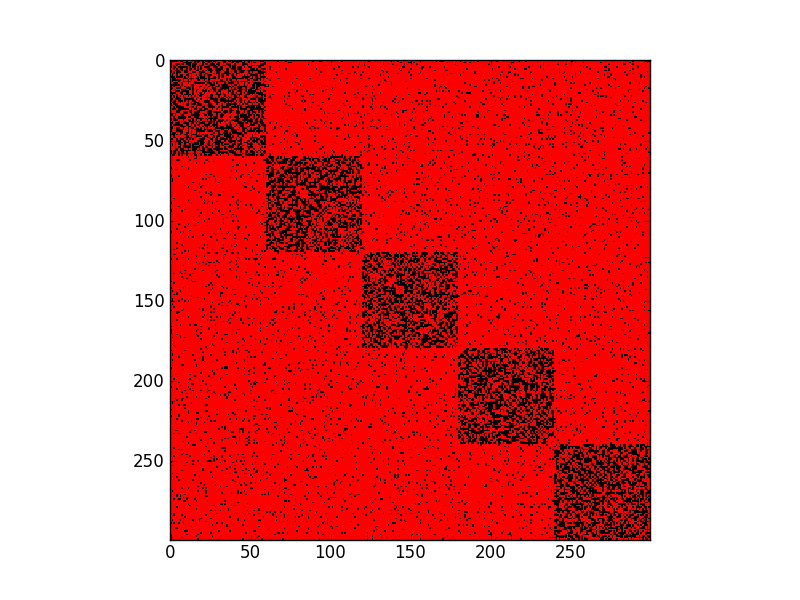}}
  \subfloat[][]{\includegraphics[width=.25\textwidth]{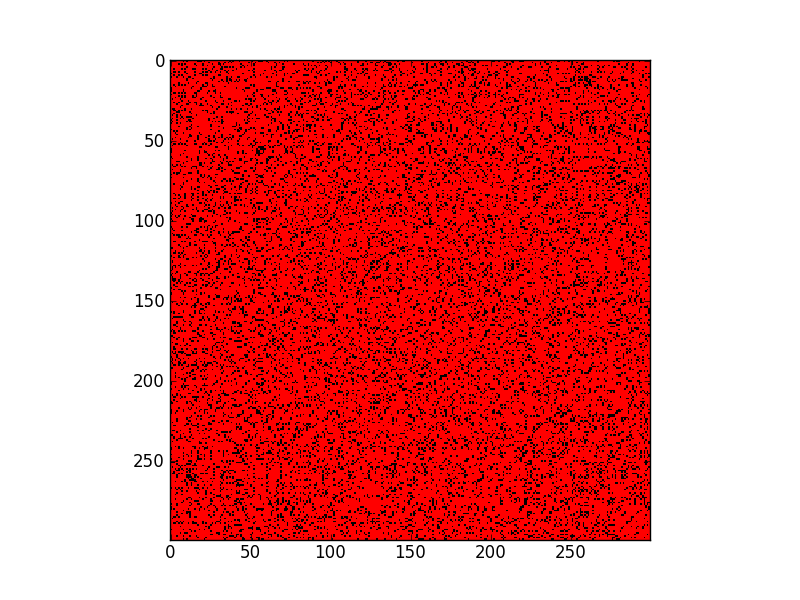}}
\newline 
\hspace*{-3cm}
  \subfloat[][]{\includegraphics[width=.25\textwidth]{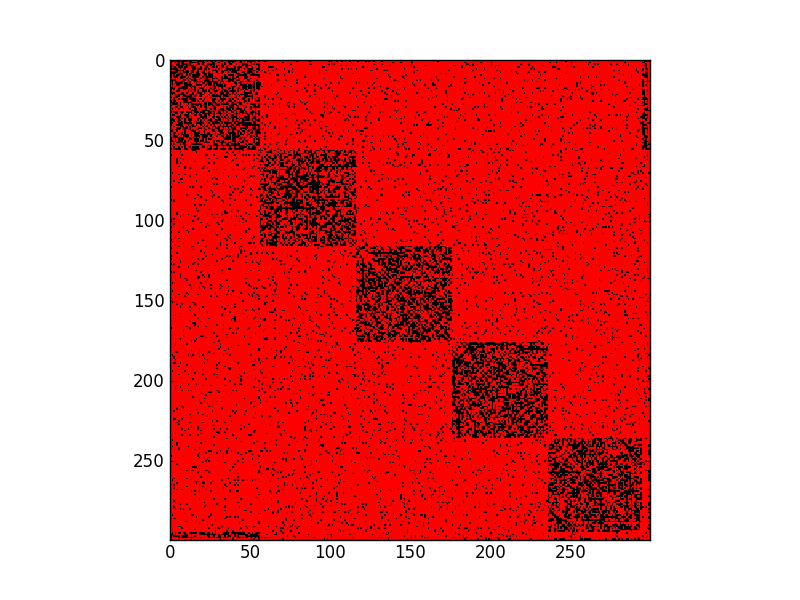}}
  \subfloat[][]{\includegraphics[width=.25\textwidth]{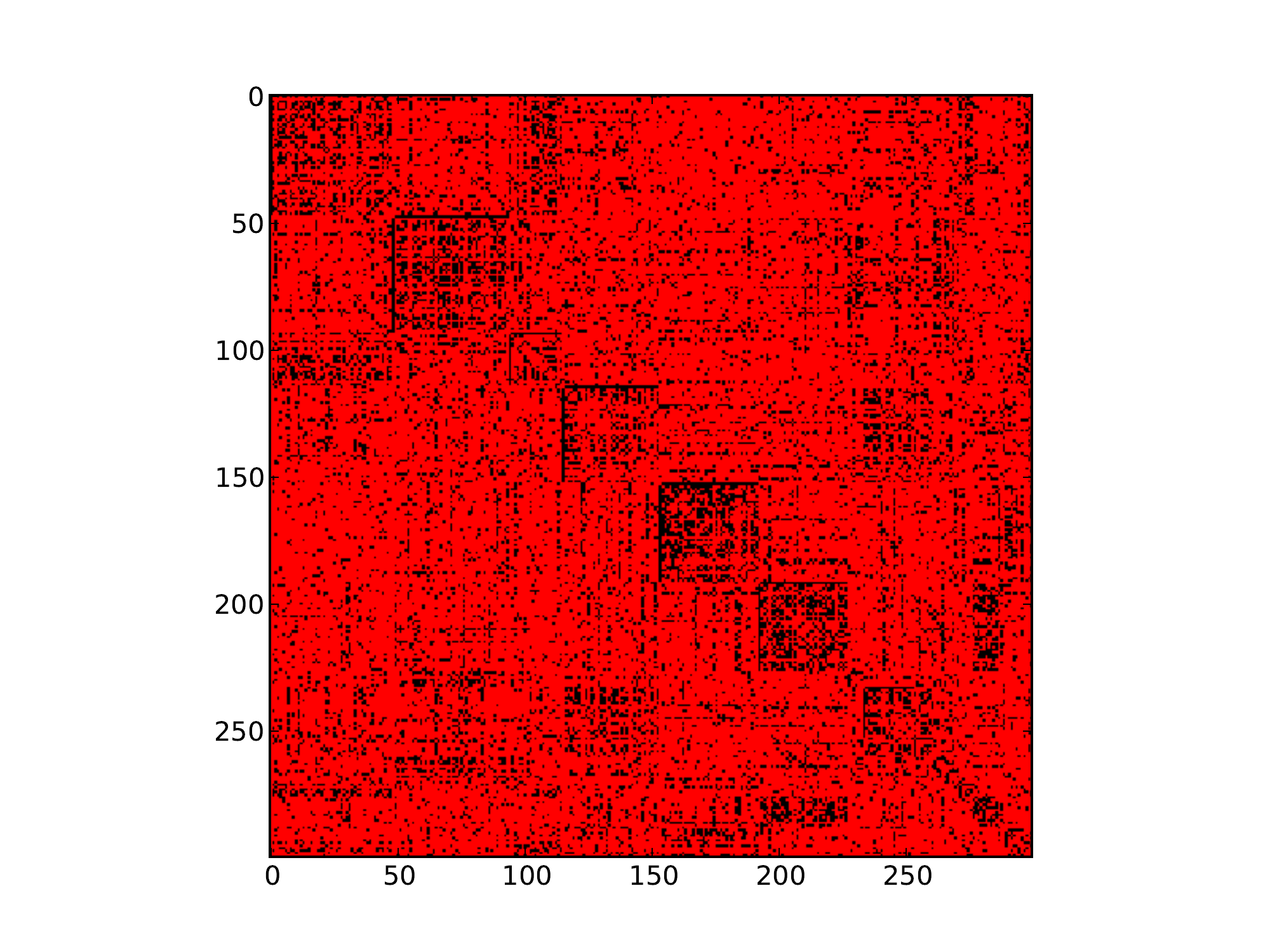}}
  \caption{(a) Matrix representation of a realization of an SBM with $n=300, k=5, p=0.55, q=0.07$. (b) Same graph with random relabeling of nodes which serves as is the input to the algorithm. (c) Communities recovered by our algorithm. (d) Communities recovered by \textsc{CC-PIVOT} algorithm.}
  \label{fig:sub1}
\end{figure}

\bRemark The values of $p$ and $q$ are designed such that the resulting random graphs have well-knit clusters. Here, the emphasize is to only compare cluster editing cost obtained by our method with those obtained by \textsc{CC-PIVOT}.
\eRemark

The actual performance for recovering communities is depicted in Fig. \ref{fig:sub1} where matrix representation of the adjacency matrix of a realization of a SBM is used to visualize community structure. As it can be seen from the figure, almost all identities are recovered correctly.

\subsection{LFR Random Graphs}
Stochastic block models are not realistic in the sense that they usually have communities with approximately the same size and furthermore, all vertices have the same degree. In realistic networks, the degree distributions are usually skewed and size of communities varies. We consider another class of random graphs introduced in \cite{lancichinetti2008benchmark} which is known as LFR benchmark graphs. In these models, the distributions of both node degrees and number of communities admit power laws.

The construction is as follows: Let $n$ be the number of nodes. First a sequence of community sizes distributed by power law with exponent $\tau_1$ is generated. The degree of each node is distributed by power law with exponent $\tau_2$. Each node within a community of size $k$, shares a fraction $1-\mu$ of his corresponding edges with the members of his community and a fraction $\mu$ of his edges with other communities. The connection of edges is done similar to Configuration Model in such a way that the degree sequence is maintained.

\begin{figure}[!t,scale=0.4]
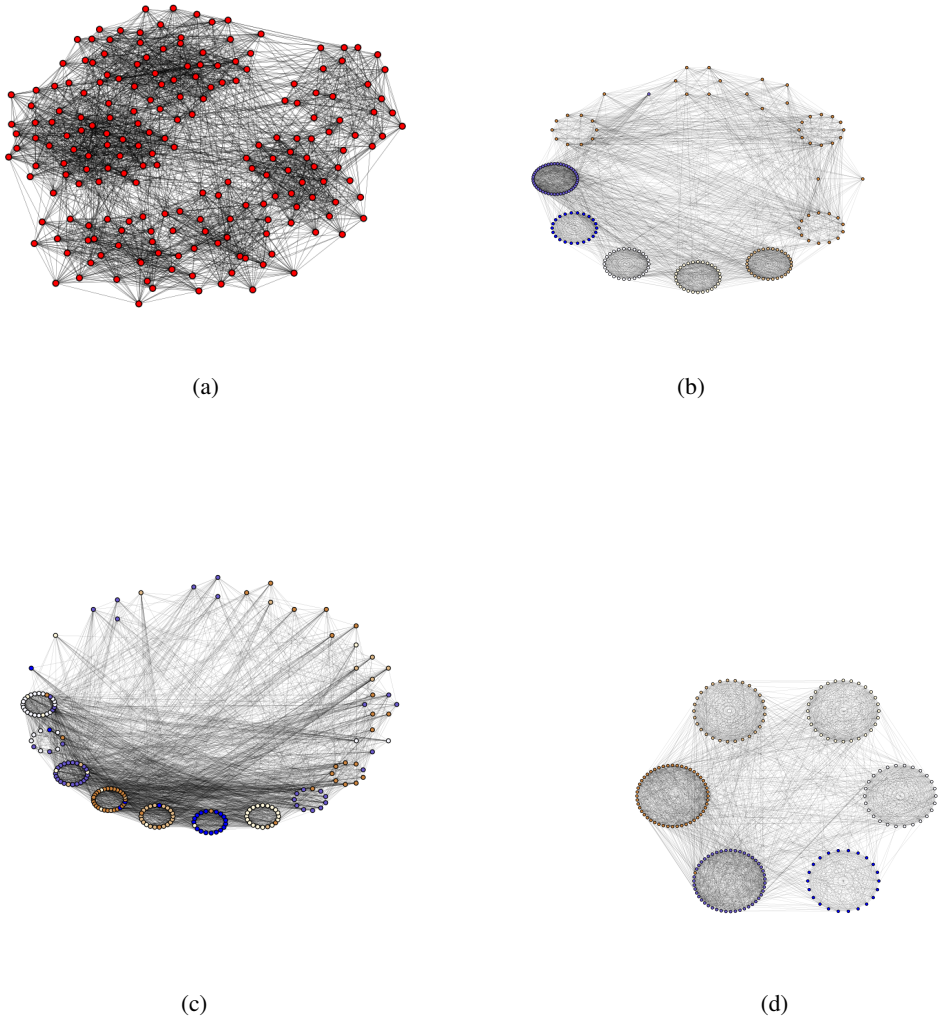

  \centering
\hspace*{-.5cm}
  \subfloat[][]{\includegraphics[width=.4\textwidth]{lfr_orig.pdf}}
\hspace*{-1cm}
  \subfloat[][]{\includegraphics[width=.4\textwidth]{lfr_ours.pdf}}
	\newline
\hspace*{-3cm}
  \subfloat[][]{ \vspace*{5cm} \includegraphics[width=.6\textwidth]{lfr_cc.pdf}}
\hspace*{-1.7cm}
  \subfloat[][]{ \includegraphics[width=.4\textwidth]{lfr_ours_final.pdf}}
  \caption{(a) An LFR random graph with $200$ nodes and $6$ communities. (b) Communities recovered by our algorithm. (c) Communities recovered by \textsc{CC-PIVOT} algorithm. (d) Communities recovered by our algorithm with fine tuning of algorithm parameters.}
  \label{fig:sub2}
\end{figure}

Fig. \ref{fig:sub2} shows a realization of LFR graph with $n=200$ nodes and $6$ communities and parameters $\tau_1=2$, $\tau_2=3$ and $\mu=0.25$ with average degree equals $30$. The communities are showed with different colors in circular layout for convenience. As it can be seen, the \textsc{CC-PIVOT} tends to create lots of singletons compared to ours. Observe that in our algorithm if we wait and track the clusters up before coalescence happens, we actually can recover communities as depicted in Fig. \ref{fig:sub2} (d).  

\subsection{Real World Network}
\begin{figure}[!h,scale=0.2]
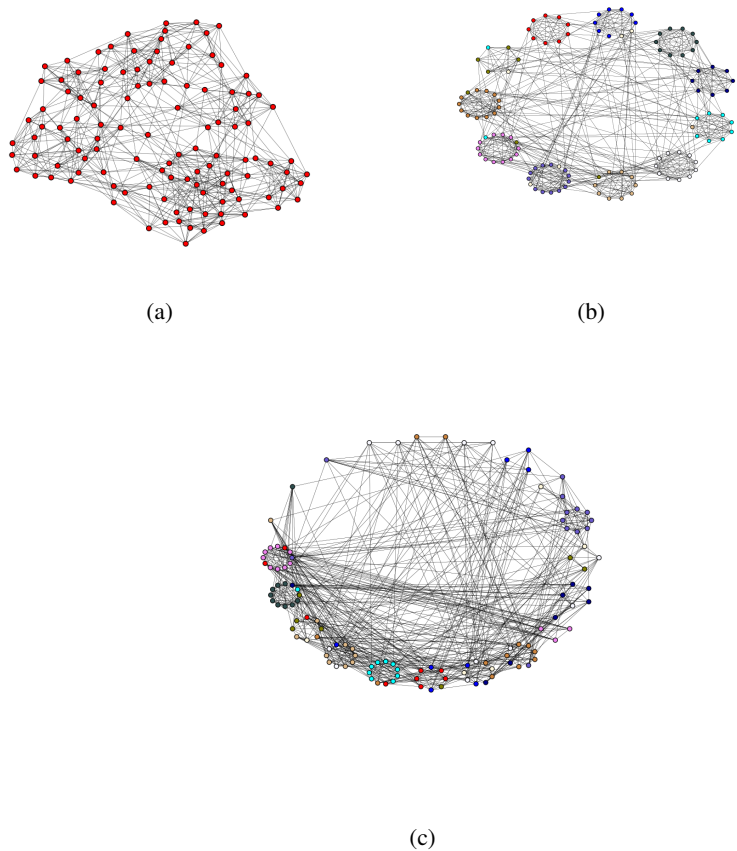

  \centering
\hspace*{-1cm}
  \subfloat[][]{\includegraphics[width=.3\textwidth]{foot_orig.pdf}}
\hspace*{-.8cm}
  \subfloat[][]{\includegraphics[width=.4\textwidth]{foot_ours.pdf}}
	\newline
\hspace*{-3cm}
  \subfloat[][]{\includegraphics[width=.5\textwidth]{foot_cc.pdf}}
  \caption{(a) American Football College benchmark network with $115$ nodes. (b) Communities recovered by our algorithm. (c) Communities recovered by \textsc{CC-PIVOT} algorithm.}
  \label{fig:sub3}
\end{figure}

 We tested the performance of the algorithm on a real world benchmark network of American football games between Division IA colleges during regular season Fall 2000 \cite{girvan2002community}. There are in total 115 teams and 615 edges in the graph. Every edge represents a game between corresponding end nodes. As depicted in Fig.~\ref{fig:sub3}, almost all nodes in the graph are assigned to their corresponding community correctly. The nodes with same color belong to the same cluster (note that the identity of nodes is not revealed to the algorithm).

\cleardoublepage
\bibliographystyle{IEEEtran}
\bibliography{Ref,univcod,cftpRef,RecSysRef}
\end{document}